\documentclass{article}
\usepackage{graphicx} 

\usepackage[affil-it]{authblk}
\usepackage{natbib}
\usepackage{color}
\usepackage{tikz}
\usetikzlibrary{shapes,decorations,arrows,calc,arrows.meta,fit,positioning}
\tikzset{
    -Latex,auto,node distance =1 cm and 1 cm,semithick,
    state/.style ={ellipse, draw, minimum width = 0.7 cm},
    point/.style = {circle, draw, inner sep=0.04cm,fill,node contents={}},
    bidirected/.style={Latex-Latex,dashed},
    el/.style = {inner sep=2pt, align=left, sloped}
}

\begin{document}

\title{Agent modelling, statistical control, and the strength of middle knowledge}
\author{Thomas Chesney, Tim Gruchman, Robert Pasley, Altricia Dawson, Stefan Gold}
\date{September 2024}

\affil{Nottingham University Business School}
\affil{University of Kassel}
\affil{Dortmund University}

\Large

\maketitle

\abstract{This methods article concerns analysing data generated from running experiments on agent based models to study industries and organisations. It demonstrates that when researchers study virtual ecologies they can and should discard statistical controls in favour of experiment controls. In the first of two illustrations we show that we can detect an effect with a fraction of the data needed for a traditional analysis, which is valuable given the computational complexity of many models. In the second we show that agent based models can provide control without introducing the biases associated with certain causal structures.}

\section{Introduction}
This methods article concerns the statistical analysis of data generated by agent-based models when used for social science, such as studying organisations and industries. Its main claim is that agent modellers should discard statistical controls in favour of experiment controls. We demonstrate this through two illustrations. The first models a shock in a supply chain and features testing for a difference between two groups (a t-test style analysis); the second models industrial accidents and features estimating the linear relationship between a dependent and independent variable while maintaining control (a multiple linear regression style analysis).

Science is a method that is used to understand, explain and predict the world. The scientific method is built on control: science is the act of comparing while exercising control. In its simplest form, two situations that are as identical to each other as possible are studied; a change is made in one; the outcomes of both are compared to answer the question: what difference did the one change make? Control is needed to ensure the situations are near identical. Broadly, there are two types of control: experiment control and statistical control, with experiment control being seen as stronger \citep{Wysocki22}\footnote{Throughout we will use the word `situation' as used above, rather than `group' or `class' or `factor' or any similar label that could be applied to mean the same thing but which might have a second meaning.}.

Agent-based modelling has a long history of use in organisation studies and industrial systems \citep{Gold20, Dijkema15, Fioretti13, Kozlowski13, Borrill11}. When used for social science an agent-based model is a representation of a target, the target being the phenomena under investigation: a team, a department, an organisation, an industry are examples. The model is made up of independent and heterogeneous objects of computer code called agents, and an environment within which they interact \citep{Wilensky15, Axtell01}. The modeller repeatedly performs experiments on their model (see \citet{Chesney21} and \citet{Winsberg10} for a justification of agent models as subjects for experimentation).

Agent modelling is important for studying industrial ecologies and has particular relevance for sustainable development. Agent models also answer recent calls for more formal theory  \citep{Vancouver20}. Implementing a theory as a computational model makes an imprecise social science theory formal and precise, and allows for rigorous testing \citep{Edwards10, Edwards10a, Adner09}. In addition, a model/theory can be explored to generate novel hypotheses, and computational theories can be tested in ways that would be expensive, unethical, or impossible otherwise \citep{Chesney21, Vancouver20, Vancouver10}.

Agent models give researchers enormous control as explained in the next section. Then the two illustrations are presented with a discussion. The Supplementary Online Materials (SOM) provide all the code used including the agent models, the test for a difference of medians, and the calculation of the regression estimates. They are available at https://github.com/ThomasCNotts/MiddleKnowledge.

 \section{Agent modelling and control}
By analogy, social scientist agent modellers can be thought of having a `God's eye view' of the societies they create--at all times they can access all values for all parameters of all agents. Variables that are exogenous to a model can be thought of being injected into it by the researcher. Two examples of such variables that will be used later are the weather and the age of employees. These are part of the model but they are not being modelled. (They could be modelled, but to model the weather for instance would require enormous computing power to process numerous complicated variables, and after all that the end result would not be what the researcher is actually interested in.) With all of the above comes a great level of control. Agent modellers can control precisely the values of parameters in a model but more than this they have access to something referred to as---to continue the God analogy---middle knowledge.

Middle knowledge is a power ascribed to God under a thesis called Molinism \citep{Craig00}. This sees God as having knowledge of counterfactuals, the ability to know what would happen in a situation even though that situation has never---or even could never---arise. Under this thesis for example, God knows what you would say to Cleopatra if you met her, even though you never will.

Agent modellers have this ability. At a point in a model run when Choice A is taken we get to see the result; but we can also rewind that same model to that same point and run it forward from there to see what would happen if Choice B had been taken instead. Note that this is subtly different from---and analytically stronger than---running the model again repeatedly with different parameter values each time. Instead two identical situations are created with only one difference, the choice of A or B.

When statistical controls (or control variables) are employed they are used to ensure that the situations being compared are as near identical as possible. Statistical controls are used to mathematically hold all else to be equal. They allow for analyses of data that have not been collected under experimental conditions \citep{Shiau24} and data that have been collected in experiments using human (and therefore not identical) participants \citep{Schram19}. For a review of control variables used in studying people at work see \citet{Bernerth16}.

It seems that statistical controls are not well understood, at least in some sub-fields. A recent information systems editorial \citep[p.2]{Shiau24} for example claims ``[m]any IS researchers feel confident that including [control variables] will lead to clean results and the discovery of `true' relationships'' and that these variables are rarely justified, especially in hypotheses. If bad controls are included in a statistics test the results become biased and misleading. There must be a theoretical reason for including a control, a researcher cannot simply `throw the kitchen sink' at the problem, see: \citet{Cinelli22, Wysocki22, Becker16, Bernerth16, Atinc12, Becker05}.

The reason statistical control is necessary in addition to experiment controls is because no two groups of individuals will ever be identical. This is not the case with two groups of agents. Middle knowledge allows us to run experiments with a heterogeneous mix of agent participants yet analyse the resulting data as if it was collected from identical situations.

\begin{figure}
\begin{tikzpicture}
    \node[state] (a) at (0,0) {$E1$};
    \node[state] (b) [below right =of a] {$b$};
    \node[state] (c) [above right =of b] {$E2$};
    \node[state] (d) [below left =of b] {$x$};
    \node[state] (e) [below right =of b] {$y$};


   \path (a) edge (b);
   \path (c) edge (b);
   \path (a) edge (d);
   \path (c) edge (e);
   \path (d) edge (e);

\end{tikzpicture}

\caption{This causal structure is known as M-bias. Controlling for $b$ statistically will spoil the relationship between $x$ and $y$.}
\label{mbias}
\end{figure}
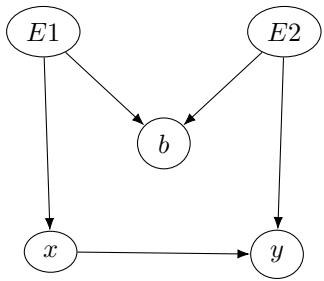

\section{Two illustrations}
We now present two illustrations, one of a test for a difference (such as a t-test) and one of a multiple linear regression. Both are in areas of industrial ecology that have previously been studied using agent models: supply chain shocks (for example: \citet{Kunz23, Rahman21}) and industrial accidents (for example: \citet{Chandra15}). In the first, a market shock causes buyers to change their opinions of the ethics of buying a product. We use middle knowledge to create a paired statistics test and contrast this with an unpaired test. The second illustration concerns an example of problems caused by the inclusion of inappropriate control variables. The causal structure studied is M-bias, which is illustrated in Figure \ref{mbias}. In the figure variable $b$ is a `bad control'. It is correlated with the independent variable of interest $x$ and the dependent variable $y$. Controlling for it will open what is referred to as a back-door path, spoiling an otherwise unbiased estimate \citep{Cinelli22,Pearl15}. A full specification of the models is provided in the SOM with enough detail given here to understand the output.

\subsection{Test for a difference}
In this model a group of buyers have a range of attitudes towards the ethics of purchasing a certain product. The buyers mix with each other and change their attitudes through a social diffusion mechanism explained below. Part way through this mixing a market shock might or might not happen. The shock can be imagined as a news outlet reporting on modern slavery in the supply chain. The model is used to study the dynamics of attitude change caused by a shock. Here we look at the number of people that end up with the same negative attitude toward the product when a shock does and does not happen.

Based on the Likert scale approach to measuring attitudes \citep{Likert32}, the model uses a simple ordinal scale. A typical Likert scale will have 5 or 7 options; here we use 5, referred to as `responses'. An individual has an `intensity' value for each response initially assigned at random. The ordinal scale refers to an individual's preference ordering of their five responses. On a Likert scale the response labels would be statements such as `Strongly agree' to `Strongly disagree' but here a wider range is possible. Here the statements could be `I do this' through to `I never do this'; or `I love this' to `I hate this'; or `I've fully adopted this' to `I've fully rejected this'. (These example labels are themselves ordinal but in fact they do not need to be--they could simply be nominal labels. An example of a nominal scale of responses might be: `I did this once', `I would do this at the weekend', `I could do this after drinking alcohol', `I would never do this'.)

A person's intensity is never compared directly with another's. This makes the concept of attitude intensity entirely subjective, as it is in nature. When two people meet, their attitude towards ethics diffuses or `rubs off' a little on each other. The model is agnostic to the psychological mechanism involved in this but the sources that were used to develop it involve ordinal thinking as previously described, for example: \citet{Aronson16, Myers82, Akers79}. The way Person 1's culture diffuses to Person 2 is:

\begin{enumerate}
\item Person 1's chosen ethical response---their response with the highest intensity---is identified, referred to as High\_Response.
\item Person 2 increases the intensity in their High\_Response by a small amount.
\end{enumerate}

The same thing then happens for Person 2's culture diffusing to Person 1. The model allows for different approaches to having two people meet. At one extreme people only meet their immediate neighbours. At the other extreme, people can meet any other person. The version used here features a probability parameter which dictates the likelihood that a person meets their direct neighbour versus meeting any other person. This is used as a `tweakable' parameter to vary between the two approaches.

A shock is an event that suddenly increases the intensity value in one of the responses (in this case the most negative one) in a randomly chosen group of susceptible agents by a parameter representing the size of the shock. The output is a count of all the agents who have that response as their High\_Response after a set time. A typical analysis strategy might be to explore the model under a range of parameter values (a sensitivity analysis), then run the model a number of times with set parameter values decided during the exploration and include a shock, run the model with the same values the same number of times and not include a shock, and compare the two for any difference using a two-group test. However, we can use middle knowledge to create a paired test, which is a stronger test \citep{Zimmerman97}. To do this the steps are:

\begin{enumerate}
\item run the simulation until the time of the shock
\item save the state of the simulation (including all agent parameter values and all values for the randomisation processes that will be used after the shock; more will be said on this in the second illustration)
\item the shock happens
\item the rest of the simulation runs and produces output
\item the simulation is reset to be exactly as it was just before the shock by loading in the saved state
\item the shock does not happen
\item the rest of the code runs and produces output
\end{enumerate}

We ran the simulation like this to generate 200 pairs (which is of course 400 output datapoints). Then we discarded middle knowledge and ran the simulation again to generate 400 datapoints for an unpaired test. For all these data the same simulation parameters were used throughout\footnote{These were: number of agents = 225, local versus global probability (the `tweakable' parameter) = 0.95 chance of meeting someone local, time of shock = time period 50, size of shock = 10, number of people shocked = 50, final time period = 100.}.

For the comparison of shock versus no shock we test for a difference of medians to ask: what is the probability of the observed difference of medians (the effect size) occurring randomly?\footnote{The reason we test for a difference of medians rather than the more usual difference of means is 1) the data are discrete and median is more appropriate and 2) to highlight that this is straightforward to do. Full details of how this test was performed along with an explanation of the code are given in the SOM. The approach used is based on \citet{Downey11}.} When the paired data are analysed, the effect size is 4 and the probability of observing this at random is $p=0.00$. For the unpaired data, the effect size is 0.5 and the probability of observing this at random is $p=1.00$. (In addition we ran a standard t-test for a difference of means. The results are shown in Table \ref{ttest}.) 

Note that the paired test was able to detect a strong relationship, the unpaired was not. In fact, even with four times as many datapoints (1,600), an unpaired test was still unable to detect the relationship (p = 0.03).

\begin{table}
\centering
\begin{tabular}{rrrr}
\hline
paired test: \\
effect size & t & df & p-value \\ 
\hline
4.61 & 12.34 & 202 & < 2.2e-16 \\
\hline
unpaired test (Welch Two Sample t-test): \\
effect size & t & df & p-value \\ 
\hline
2.60 & -0.80 & 381.94 & 0.42 \\
\hline
\end{tabular}
\caption{A t-test for a difference of means using the shock model data.}
\label{ttest}
\end{table}

\subsection{Controlling variables during a model run}
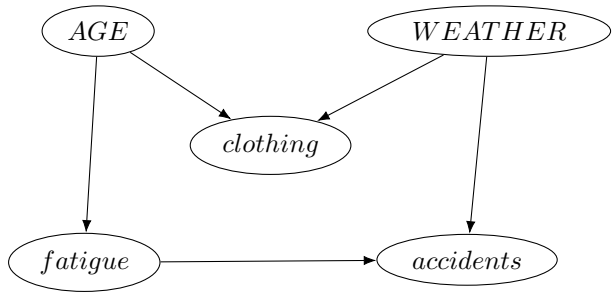
\begin{figure}
\begin{tikzpicture}
    \node[state] (a) at (0,0) {$AGE$};
    \node[state] (b) [below right =of a] {$clothing$};
    \node[state] (c) [above right =of b] {$WEATHER$};
    \node[state] (d) [below left =of b] {$fatigue$};
    \node[state] (e) [below right =of b] {$accidents$};


   \path (a) edge (b);
   \path (c) edge (b);
   \path (a) edge (d);
   \path (c) edge (e);
   \path (d) edge (e);
\end{tikzpicture}

\caption{An example structure featuring M-bias: industrial accidents in a factory. The exogenous variables influence agent fatigue, the number of accidents and the amount of clothing worn. The relationship of interest is between fatigue and accidents.}
\label{accidents}
\end{figure}

In this second illustration we show how an agent model can control for all variables in a situation that features potentially bad controls without biasing the results, something that cannot be guaranteed using statistical control \citep{Ding15, Pearl10}. In this case the bad control, if included in statistical analysis, would form an M-bias structure \citep{Griffith20} but the basic argument applies to other structures. The causal model is shown in Figure \ref{accidents}. The situation being modelled is industrial accidents in a factory. Accidents are measured at the individual level--the dependent variable is the number of accidents caused by an employee agent. The model features two exogenous variables, elements that are part of the model but are not being modelled: agent age and the weather. The independent variable of interest is agent fatigue. Older agents tend to feel more fatigue than younger agents. There tend to be more accidents when it is cold. Also included is a measure of the weight of clothing worn by agents---when it is cold and when agents are older they tend to wear warm clothes which are heavier. The factory has an outside yard (which is why weather can impact accidents).

The model runs as follows. Firstly each area in the factory is randomly assigned a danger value and the weather is randomly set. The agents are then instantiated and each given a random age. Weather and age are used to set the amount of clothing each agent wears. Agent fatigue is set according to their age and a random element. Then the simulation starts. In each period, the agents move about the factory and may have an accident caused by the weather (e.g. slipping on ice or sliding in the rain) or because they are in a more dangerous area. Certain areas in the factory are more dangerous than others. Using an approached based on \citet{Palaniappan07}, if an agent's fatigue is greater than an area's danger value, there is a probability proportion to their fatigue that they will have an accident. Fatigue increases in each time period but is reduced again back to a baseline after an accident.

As before we use middle knowledge, although this time it may be less obvious as to why because the simulation state is saved (Step 2 from the previous illustration) after model instantiation (which is usually called `setup' by agent modellers) but before the first time period. Setup includes setting global variables (like the weather), creating all agents (which in this case means assigning them an age and modelling their fatigue), and creating all random numbers that are about to be used. This randomisation, which was mentioned in Step 2 in the previous illustration but not explained, is both vital and well understood by agent modellers. For any readers less familiar, agents do not act simultaneously (this is the major difference between agent models and cellular automata like Game of Life \citep{Bays10}). Instead agents take turns when acting; they take turns in a random order, and each time they act this order is refreshed randomly so that no agent always gets to go first (or second, third, or last, or anything else). If this is not done, the results will be biased (see \citet{Huberman93} for an example and discussion). Contrary to popular belief computers cannot create truly random numbers, so any list of random numbers that starts from the same computer state will be identical. The upshot of this is that to achieve middle knowledge we must make sure that this state is saved so that all orders and other randomisations will be identical for pairs of simulation runs. How to actually do this varies greatly depending on the software platform being used (see the SOM for one approach). However it is done, we seek to hold all else equal and increase fatigue by one unit to determine the effect this has on accidents. Traditionally this might be achieved after data collection using a linear model such as:

$accidents = coef_1 * fatigue + coef_2 * clothing + error$

with the coefficients $coef_1$ and $coef_2$ measuring the relationship between the independent variables and accidents, specifically the amount the independent variable must be multiplied by to determine the change in the dependent from a change of one unit in the independent variable, while holding other variables in the model constant.

Firstly we ran the model without middle knowledge 160 times and Table \ref{probs} shows the results of a linear regression when controlling for clothing (top panel) and not controlling (lower panel). The table highlights the problem of including a bad control: the coefficient for fatigue has changed considerably, as has the p-value.

We then ran the model using middle knowledge to create 80 pairs (which is 160 output datapoints) as illustrated in Table \ref{fat}. The steps to do this are:

\begin{enumerate}
\item run setup
\item save the model state
\item run the model
\item reset the state
\item increase each agent's fatigue by one unit
\item run the model
\end{enumerate}

These data were used to produce the same output as a multiple regression--a regression coefficient, an intercept and a p-value. this was done while controlling for clothing, a bad control. To calculate the coefficient we simply produced the difference of the mean of $accidents_1$ and $accidents_2$; the standard deviation of $accidents_1$ - $accidents_2$ gives the standard error and the p-value comes from a correlation test--see Table \ref{fat} for the details and the SOM for the code\footnote{Indeed, we could very easily report the median difference as before instead of the mean---in fact it would be 9---but this would not be the traditional regression output.}. The results are: intercept estimate = -52.80, coefficient estimate for fatigue = 2.56, p-value = < 2.2e-16, standard error = 10.88. There is no estimate for clothing as one is not needed--clothing was controlled using experiment control.


The coefficient estimate for fatigue is 2.56 which we feel is a more accurate estimate than that produced by the linear regression in the lower panel of Table \ref{probs}, for the reasons laid out throughout this paper. Even if readers are skeptical of this claim, the fact is that the coefficient was derived while controlling for clothing as within each data pair the clothing of each agent is identical. When this was done using statistical controls (upper panel of Table \ref{probs}) a widely inaccurate estimate was created.

\begin{table}
\centering
\begin{tabular}{rrrrr}
  \hline
agent & fatigue & fatigue + 1 & accidents\textsubscript{1} & accidents\textsubscript{2} \\ 
  \hline
1 & 5.2 & 6.2 & 80 & 82\\
2 & 3.9 & 4.9 & 73 & 78\\
3 & 9.8 & 10.8 & 56 & 57\\
4 & ...\\
\hline
\end{tabular}
\caption{Example output from the accidents model. Each agent is assigned a fatigue and the model is run to produce $accidents_1$. Then the model is reset using middle knowledge. Then fatigue is increased by one unit and the model run again to produce $accidents_2$. From this the coefficient is produced by $absolute(mean(accidents_1) - mean(accidents_2$)), the standard error by $standard.deviation(accidents_1 - accidents_2)$, and the p-value by a Pearson's product-moment correlation test, $correlation.test(fatigue, accidents)$.}
\label{fat}
\end{table}

\begin{table}[ht]
\centering
\begin{tabular}{rrrrr}
  \hline
 & estimate & std. error & t-value & p-value \\ 
  \hline
  intercept & 181.26 & 222.96 & 0.81 & 0.42 \\ 
  fatigue & 14.13 & 12.94 & 1.09 & 0.28 \\ 
  clothing & -17.14 & 18.48 & -0.93 & 0.36 \\ 
   \hline
\end{tabular}
\begin{tabular}{rrrrr}
  \hline
 & estimate & std. error & t-value & p-value \\ 
  \hline
  intercept & -25.45 & 2.42 & -10.51 & 0.00 \\ 
  fatigue & 2.14 & 0.04 & 58.57 & 0.00 \\ 
   \hline
\end{tabular}
\caption{The results of linear regression analysis when controlling for clothing and fatigue. The Upper panel shows the estimates when using statistical control, the lower panel when using middle knowledge. In the lower panel, the estimate for fatigue has changed size considerably and is now significant.}
\label{probs}
\end{table}

\section{Discussion}
Agent modelling is an important research method for organisation studies allowing for experiments that would be expensive or unethical otherwise, and giving researchers the opportunity to implement and explore formal theory. This paper contributes to a discussion of how the such models should be run and analysed to best maintain control. By using middle knowledge, data pairs can be created that are identical in all aspects other that the treatment under consideration. If more than one independent variable is of interest then the identical model can be run more than twice to create not pairs but a related set. Each set would become a row in the results, creating a dataset where data within each row come from an identical model, but between rows the models would be distinct (by having different initial parameter values). Columns in such a dataset could be analysed to extract the relationships between the independent variables and the dependent. By analysing agent models in this way plays to one of their major strengths.

Middle knowledge may be generated by altering strategies or decisions of single actors while keeping all other model parameters constant. More sophisticated approaches pursue configurational analyses that shed light on interactive effects since effects often do not materialize by change of single factors operating in isolation but bundles of factors (some of which may be conceived as context variables) (e.g., \citet{Ketchen22}. A context-sensitive research approach is important to generate theories that can guide strategic change towards an ecologically sustainable and socially equitable development \citep{Johns06}. It is the particular strength of agent modelling to comprehend a vast variety of context variables and run simulation-based controlled experiments \citep{Wilensky15} that help predicting the impact of certain practices, strategies or policies, or bundles of those. This enables knowledge about potential situations--which have not (yet) materialized and might never materialize, i.e. middle knowledge--from which real-world managerial and policy recommendations can effectively be derived. Simulation leverages its principal strength when engaging in systematic and “creative experimentation to produce novel theory” \citep[p.480]{Davis07}, i.e. explaining and predicting events or actions and their boundary conditions \citep{Wacker98}.

Middle knowledge is of particular relevance for the research fields of sustainable development, sustainability management and ethics. In the field of climate research, for example, the simulation of the effects of certain concentrations of CO$_2$ equivalents in the atmosphere on ecosystems, crop yields and human settlements, even before the situation has occurred (e.g., \citet{Still99}, is extremely helpful for informed action. Strategic responses to societal grand challenges \citep{George16} such as educational poverty, insufficient health care or modern slavery requires business, civil society and public authorities to gain a deep understanding of the situation that causes persistency of those “wicked” problems \citep{Rittel73}. The impact of specific interventions on systemic behaviour and key outcome parameters has to be analysed in comparison to baseline scenarios \citep{Oliva19}, as do potential (subsequent) behavioural or strategic adaptations of key stakeholders \citep{Atasu12}. The effectiveness of public policy in promoting sustainable production and consumption patterns on a societal scale depends crucially on the interplay between design and implementation of regulation and the responses of key stakeholders \citep{Bodrozic24}. Controlled experimentation with strategic interventions and adaptive responses by various actor groups are needed to devise effective tools for keeping society and economy within planetary boundaries and on strong social foundations, while avoiding undesired side effects that such interventions may generate \citep{Carter20, Merton36}. 

Humanity lives in times of crisis, reflecting multiple dimensions of simultaneous environmental (e.g., climate warming, loss of biodiversity) and social issues (e.g., poverty, inequalities, political instability and wars) (e.g., \citet{Gold24}. At the same time, this crisis offers an unprecedented window of opportunity to transform societies and economies worldwide, and to transition towards the path of sustainable development \citep{Bodrozic18}. The great task of humanity in the current era of crisis is to allow this transition to be `by design' and not `by disaster'. This requires substantial knowledge about the current situation at hand, and the potential situations that might (or might not) unfold during the non-steady-state period of transition. Such knowledge includes future states and events, likelihood of occurrence, boundary conditions, and the systemic impact of action by the various stakeholders such as consumers, public authorities, business and non-governmental organisations. Indeed, creating middle knowledge through controlled experiments, by tools such as agent modelling, is the scholarly contribution to a sustainability transition `by design'. Such knowledge helps humanity build its capabilities to shape the transition, navigate its paradoxes and intricacies \citep{Matthews24}, and ultimately, in words of the United Nations' Sustainable Development Goals, `leaving no one behind'.

For the purpose of open access, the authors have applied a Creative Commons Licence to any Author Accepted Manuscript version arising.

\bibliographystyle{plainnat}
\bibliography{Chesney}

\end{document}